\input lanlmac
\input epsf.tex
\input mssymb.tex
\overfullrule=0pt

\newcount\figno
\figno=0
\def\fig#1#2#3{
\par\begingroup\parindent=0pt\leftskip=1cm\rightskip=1cm\parindent=0pt
\baselineskip=11pt
\global\advance\figno by 1
\midinsert
\epsfxsize=#3
\centerline{\epsfbox{#2}}
\vskip 12pt
{\bf Fig.\ \the\figno:} #1\par
\endinsert\endgroup\par
}
\def\figlabel#1{\xdef#1{\the\figno}%
\writedef{#1\leftbracket \the\figno}%
}
\def\omit#1{}

\def\IR{\relax{\rm I\kern-.18em R}}
 
\def\IC{\relax{\rm C\kern-1.4ex I\ }}

\def\pre#1{{\tt
#1}}

\def\Rc{{\check R}}

\def\der{\partial}

\def\ket#1{\left| #1 \right>}

\def\tvi{\vrule height 12pt depth 6pt width 0pt}
\def\tv{\tvi\vrule}

\def\qed{\nobreak\hfill\vbox{\hrule height.4pt%
\hbox{\vrule width.4pt height3pt \kern3pt\vrule width.4pt}\hrule height.4pt}\medskip\goodbreak}
%
%
\lref\RS{A.V. Razumov and Yu.G. Stroganov, 
{\sl Combinatorial nature
of ground state vector of $O(1)$ loop model},
{\it Theor. Math. Phys.} 
{\bf 138} (2004) 333-337; {\it Teor. Mat. Fiz.} 138 (2004) 395-400, \pre{math.CO/0104216}.}
\lref\BdGN{M.T. Batchelor, J. de Gier and B. Nienhuis,
{\sl The quantum symmetric XXZ chain at $\Delta=-1/2$, alternating sign matrices and 
plane partitions},
{\it J. Phys.} A34 (2001) L265--L270,
\pre{cond-mat/0101385}.}
\lref\DFZJ{P.~Di Francesco and P.~Zinn-Justin, {\sl Around the Razumov--Stroganov conjecture:
proof of a multi-parameter sum rule}, {\it E. J. Combin.} 12 (1) (2005), R6,
\pre{math-ph/0410061}.}
\lref\DFZJb{P.~Di Francesco and P.~Zinn-Justin, {\sl Inhomogeneous model of crossing loops
and multidegrees of some algebraic varieties}, to appear in {\it Commun. Math. Phys} (2005),$\ \ \ \ \ \ \ $
\pre{math-ph/0412031}.}
\lref\KZJ{A. Knutson and P. Zinn-Justin, 
{\sl A scheme related to the Brauer loop model}, \pre{math.AG/0503224}.}
\lref\Pas{V.~Pasquier, {\sl Quantum incompressibility and Razumov Stroganov type conjectures},
\pre{cond-mat/0506075}.}
\lref\Jo{A. Joseph, {\sl On the variety of a highest weight module}, {\it J. Algebra} 88 (1) (1984), 238--278.}
\lref\Ro{W. Rossmann,
{\sl Equivariant multiplicities on complex varieties.
  Orbites unipotentes et repr\'esentations, III},
  {\it Ast\'erisque} No. 173--174, (1989), 11, 313--330.}
\lref\Ho{R.~Hotta, {\sl On Joseph's construction of Weyl group representations}, Tohoku Math. J. Vol. 36
(1984), 49--74.}
%
%
\lref\FR{I.B.~Frenkel and N.~Reshetikhin, {\sl Quantum affine Algebras and Holonomic Difference Equations},
{\it Commun. Math. Phys.} 146 (1992), 1--60.}
\lref\JM{M.~Jimbo and T.~Miwa, {\it Algebraic analysis of Solvable Lattice Models}, 
CBMS Regional Conference Series in Mathematics vol. 85, American Mathematical Society, Providence, 1995.}
\lref\DFKZJ{P.~Di Francesco, A. Knutson and P. Zinn-Justin, work in progress.}
\lref\SKLY{E. Sklyanin, {\sl Boundary conditions for integrable quantum systems},
J. Phys. A: Mat. Gen. {\bf 21} (1988) 2375-2389.}
\lref\DFOP{P. Di Francesco, {\sl Inhomogeneous loop models with open boundaries},
J. Phys. {A}: Math. Gen. {\bf 38} (2005) 6091-6120, \pre{math-ph/0504032}.}
\lref\LETDZJ{P.~Di Francesco and P. Zinn-Justin, {\sl Quantum Knizhnik-Zamolodchikov equation,
generalized Razumov-Stroganov sum rules and extended Joseph polynomials}, preprint SPhT-T05/130 (2005),
\pre{math-ph/0508059}.}
\lref\MEAN{P.~Di Francesco, {\sl SU(N) meander determinants}, J. Math. Phys. {\bf 38} (1997) 5905-5943,
\pre{hep-th/9702181}.}
\lref\KUP{G. Kuperberg, {\sl Symmetry classes of alternating sign matrices under one roof},
Ann. of Math. (2) {\bf 156} No. 3 (2002) 835-866, \pre{math.CO/0008184}.}
\lref\OKA{S. Okada, {\sl Enumeration of symmetry classes of alternating sign matrices and characters
of classical groups}, 
\pre{math.CO/0408234}.}
\lref\IK{A. Izergin, {\sl Partition function of the six-vertex
model in a finite volume}, Sov. Phys. Dokl. {\bf 32} (1987) 878-879;
V. Korepin, {\sl Calculation of norms of Bethe wave functions},
Comm. Math. Phys. {\bf 86} (1982) 391-418.}
\lref\PRdGN{P. A. Pearce, V. Rittenberg, J. de Gier and B.~Nienhuis, 
{\sl Temperley--Lieb Stochastic Processes},
{\it J. Phys.} A35 (2002), L661--L668,
\pre{math-ph/0209017}.}
\lref\FULHA{W. Fulton and J. Harris, {\sl Representation theory}, Springer-Verlag, New York (1991).}
\lref\JMW{M. Jimbo, R. Kedem, H. Konno, T. Miwa and R. Weston,
{\sl Difference equations in spin chains with a boundary}, Nucl. Phys. {\bf B448[FS]} (1995) 429-456,
\pre{hep-th/9502060}.}
\lref\KQ{T. Kojima and Y. Quano, {\sl Difference equations for the higher rank XXZ model with a 
boundary}, Int. Jour. of Mod. Phys. {\bf A15} (2000) 3699-3716, \pre{nlin.SI/0001038}.}
\Title{SPhT-T05/135}
{\vbox{
\centerline{Boundary qKZ equation and generalized}
\medskip
\centerline{Razumov-Stroganov sum rules for open IRF models}
}}
\bigskip\bigskip
\centerline{P.~Di~Francesco} 
\medskip
\centerline{\it  Service de Physique Th\'eorique de Saclay,}
\centerline{\it CEA/DSM/SPhT, URA 2306 du CNRS,}
\centerline{\it F-91191 Gif sur Yvette Cedex, France}
\bigskip
\vskip0.5cm
\noindent
We find higher rank generalizations of the Razumov--Stroganov sum rules 
at $q=-e^{i\pi\over k+1}$ for 
$A_{k-1}$ models with open boundaries, by constructing polynomial solutions of 
level one boundary quantum Knizhnik--Zamolodchikov equations for $U_q(\frak{sl}(k))$. 
The result takes the form of a character of the symplectic group, that
leads to a generalization of the number of vertically symmetric alternating 
sign matrices. We also investigate the other combinatorial point $q=-1$,
presumably related to the geometry of nilpotent matrix varieties.

\bigskip


\Date{09/2005}
%
%
\newsec{Introduction}

Among the many beautiful connections between statistical physics and combinatorics,
those involving integrable lattice models are probably the deepest ones. 
A major advance on the enumeration of various symmetry classes of alternating sign matrices
(ASM) was accomplished by Kuperberg \KUP\ by relating these notoriously difficult combinatorial problems
to partition functions of the  six-vertex model on a square grid with various
boundary conditions. Alternatively, Okada \OKA\ reinterpreted these partition functions as characters
of various classical Lie groups.

In this note, we address a related combinatorial mystery,
the Razumov--Stroganov conjecture \RS, that identifies the suitably normalized groundstate vector
entries of the O(1) loop model on a semi-infinite cylinder of perimeter $2n$ to
numbers of configurations of the fully-packed loop model on an  $n\times n$ square grid. A weaker
version of this conjecture, the Razumov-Stroganov sum rule \BdGN\ identifies the sum of these entries with the
number $A_n$ of ASMs of size $n\times n$, and has been proved in \DFZJ, in an inhomogeneous
version (incorporating position-dependent spectral parameters $z_i$) making extensive use of the 
integrability of the loop model. It allows for the identification between
the sum of entries of the fully inhomogeneous O(1) loop model and the very same partition function as that
used by Kuperberg for the enumeration of ASMs, the so-called Izergin--Korepin (IK) determinant \IK.
This proof was adapted to the case of the inhomogeneous O(1) crossing loop model as well \DFZJb, 
and allowed to identify the entries of the groundstate vector with the multidegrees of some matrix 
varieties \KZJ.  The case of open boundaries for the loop models was also treated in \DFOP\ both for
non-crossing and crossing loop models respectively, giving rise to other sum rules. In particular,
this established the open Razumov--Stroganov sum rule, identifying the sum of suitably
normalized groundstate vector
entries of the O(1) loop model on a semi-infinite strip of width $2n$ with the number $A_V(2n+1)$
of vertically symmetric alternating sign matrices (VSASM) of size $(2n+1)\times (2n+1)$.

In the context of $q$-deformations of quantum hall effect wave functions \Pas,
Pasquier recovered the Razumov--Stroganov sum rule (for $q=-e^{i\pi\over 3}$)
by finding the minimal
polynomial solution of the quantum Knizhnik--Zamolodchikov (qKZ) equation \FR\ for $U_q(\frak{sl}(2))$
at level $1$. In \LETDZJ, solving qKZ equations for higher rank algebras
$U_q(\frak{sl}(k))$ at level $1$ led to a nice generalization of the Razumov--Stroganov sum rule
for $q=-e^{i\pi\over k+1}$,
identifying a weighted sum of entries of the groundstate vector of the inhomogeneous $A_{k-1}$ 
interaction-round-a-face (IRF) model on a semi-infinite cylinder of perimeter $kn$ with
particular Schur functions generalizing the IK determinant and Okada's result. In particular,
in the homogeneous limit where all $z_i\to 1$, the sum rule was found to be proportional
to the number: 
\eqn\genasmnum{ A_n^{(k)}=\prod_{j=0}^{n-1}{j!\prod_{i=1}^{k-1}\big((k+1)j+i\big)!\over 
\prod_{i=0}^{k-1}(kj+i)! }}
reducing to the number of ASMs for $k=2$.

In the present paper, we extend the results of \LETDZJ\ to the case of $A_{k-1}$ IRF models
with open boundaries. To this purpose, we introduce in Sect.2 a boundary version of the 
level $1$ qKZ equation 
which incorporates a parameter $r$, playing the role of a boundary magnetic field, and such that
the ordinary qKZ equation is recovered when $r\to 0$. This extends the boundary qKZ
equation of \JMW\ and \KQ, defined respectively for $U_q(\frak{sl}(2))$ and $U_q(\frak{sl}(k>2))$,
both at level $0$. We then
use the path representation of \LETDZJ\ for the quotients of the Hecke algebra associated
to the $A_{k-1}$ models to find minimal polynomial solutions to these boundary qKZ equations in Sect.3.
At the generalized Razumov--Stroganov point $q=-e^{i\pi\over k+1}$, we compute
a weighted sum of entries of this polynomial solution, identified with the groundstate
entries of the corresponding inhomogeneous $A_{k-1}$ IRF model on a semi-infinite strip
of width $kn$, with open boundaries (Sect.4.1). This sum rule produces a simple generalization
of the Schur functions of \LETDZJ, in a form very similar to the character formulas
of Okada for symplectic groups. 
In the homogeneous limit where all $z_i\to 1$, all these polynomials collapse to polynomials
of $r$ with integer coefficients, and lead at $r=1$ to a natural generalization of the number of VSASMs:
\eqn\genvsasmnum{A_V^{(k)}(n)=\prod_{j=1}^{n-1} {(2j)!(2nk+2j-1)!\over ((k+1)j)!((k+1)(j+n)-1)!}}
which reduces to $A_V(2n+1)$ for $k=2$.

We also investigate the other ``combinatorial point" $q=-1$, corresponding to the rational
limit of the underlying integrable model (Sect.4.2). There, in the homogeneous limit,
we find that all entries of our solution tend to integers, and discuss their possible interpretation as
degrees of nilpotent matrix varieties.

\newsec{Quantum Knizhnik-Zamolodchikov equations without and with a boundary}

\subsec{$R$ matrix and path representation}

Throughout this note, we shall use the standard abstract trigonometric solution of the 
Yang-Baxter equation which reads
\eqn\trigosol{ \Rc_{i,i+1}(z,w)={q^{-1}z -q w\over q^{-1} w-q z}I +{z-w\over q^{-1} w-q z}e_i }
where the $e_i$, $i=1,2,...,N-1$ are the  generators of the Hecke algebra $H_N(\tau)$, namely subject
to the relations
\eqn\relahec{ e_ie_j=e_je_i,\ |i-j|>1, \qquad e_ie_{i+1}e_i-e_i=e_{i+1}e_ie_{i+1}-e_{i+1},
\qquad e_i^2=\tau e_i}
with the parametrization
\eqn\taketau{\tau=-(q+q^{-1}) }
A standard pictorial representation for $\Rc$ reads
\eqn\stanpict{ \Rc_{i,i+1}(z,w)=\epsfxsize=1.5cm\vcenter{\hbox{\epsfbox{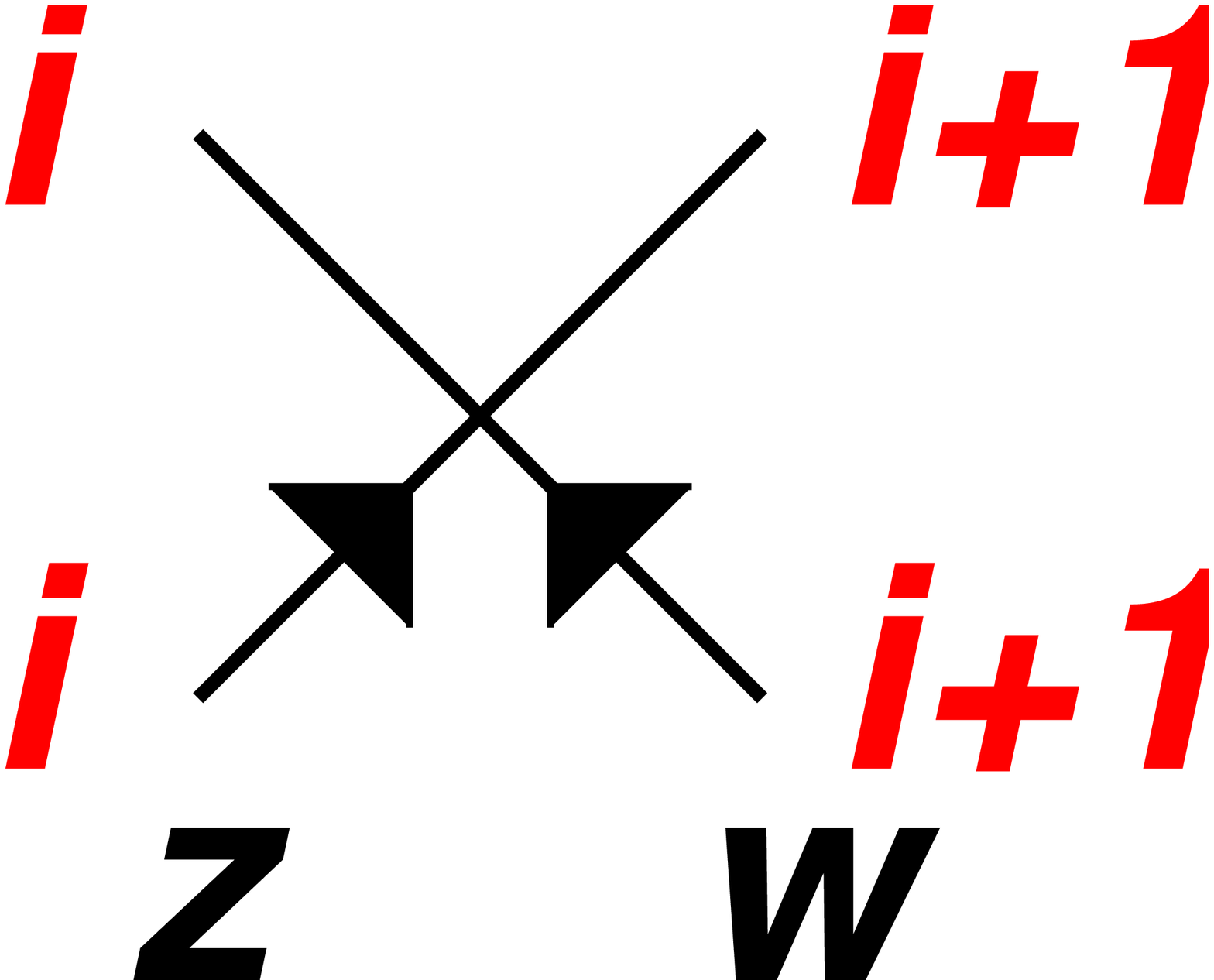}}}}
where we have also indicated the action on the spaces $i$ and $i+1$.

More precisely, we let $\Rc$ act on the so-called path representation of the $A_{k-1}$ IRF models.
To this purpose, the $e_i$ are further constrained to generate the $U_q(\frak{sl}(k))$
(also called $A_{k-1}$ or $SU(k)$) quotient $H_N^{(k)}(\tau)$
of $H_N(\tau)$ obtained by imposing the extra vanishing conditions $Y_k(e_i,e_{i+1},\ldots,e_{i+k-1})=0$,
$i=1,2,...,N-k$,
where the Young $q$-symmetrizers $Y_k$ are defined inductively by $Y_1(e_i)=e_i$ and 
$Y_{m+1}(e_i,\ldots,e_{i+m})=Y_m(e_i,\ldots,e_{i+m-1})(e_{i+m}-\mu_m)Y_m(e_i,\ldots,e_{i+m-1})$,
where $\mu_m=U_{m-1}(\tau)/U_m(\tau)$, $U$ the Chebyshev polynomials subject to
$U_m(2\cos x)=\sin (m+1)x/\sin x$. Note that $k=2$ corresponds to the Temperley-Lieb algebra $TL_N(\tau)$.
The path representation is only briefly sketched here
and will be described in detail in \DFKZJ.
The representation basis is indexed by closed paths of length $N=nk$, from and to the origin, on the oriented-link 
Weyl chamber of $SU(k)$,
generated by the $k$ vectors $u_1=\omega_1$, $u_2=\omega_2-\omega_1$, ..., $u_{k-1}=\omega_{k-1}-\omega_{k-2}$, 
and $u_k=-\omega_{k-1}$, in terms of the fundamental weights $\omega_i$. The paths are allowed to visit
only points $\lambda=\sum\lambda_i\omega_i$ with all $\lambda_i\geq 0$, and may be indexed
by their sequence of steps, and by a slight abuse of notation we write $\pi=(i_1i_2\ldots i_N)$
for the sequence $u_{i_1},u_{i_2},\ldots,u_{i_N}$; we also note $\pi_m=i_m$, 
the $m$-th step of the path $\pi$. For instance,
the path of length $nk$
closest to the origin is $(12\ldots k)^n$, namely $n$ repetitions of the sequence $1,2,\ldots,k$,
and we denote it by $\pi_f$.
Likewise, the path farthest from the origin is $(1)^n(2)^n\ldots(k)^n$, and we denote it by $\pi_0$.
A useful notation consists in representing each step $j$ by a unit segment forming an angle of 
${\pi(k+2-2j)\over 2(k+1)}$ with the horizontal direction: each $\pi$ becomes a broken line touching
the $x$ axis at its ends and staying above it. There are exactly 
$c_{nk}^{(k)}=(kn)!\prod_{0\leq j\leq k-1} j!/(n+j)!$
such paths. The number $c_{nk}^{(k)}$ is the dimension of the path representation and generalizes
the Catalan numbers $c_n=c_{2n}^{(2)}$.
The next step consists in decomposing each path in the broken line representation into
``tiles" made of (possibly glued) lozenges $L_{i,j}$ with edges corresponding to pairs $1\leq i<j\leq k$
of steps. To each tile we then associate a specific polynomial of the $e$'s. For instance, each
single lozenge with steps at positions $i,i+1$ along the paths corresponds to $e_i$. The tile decomposition
of each path is then unique up to elementary moves between tiles allowed by the algebra relations,
and we may therefore associate a unique element of $H_N^{(k)}$ to each path $\pi$, thus
forming a representation basis $|\pi\rangle$. Actually the map is onto the left
ideal $H_N^{(k)}\Omega$, 
$\Omega=Y_{k-1}(e_1,...,e_{k-1})Y_{k-1}(e_{k+1},...,e_{2k-1})...Y_{k-1}(e_{(n-1)k+1},...,e_{nk-1})$,
isomorphic to $H_N^{(k)}$.
For illustration, let us detail the case $k=3$, $n=2$. There are five paths, respectively decomposed 
as follows: defining $Y_{i,i+1}\equiv Y_2(e_i,e_{i+1})=e_ie_{i+1}e_i-e_i$, $Z_{i,j}=e_ie_j-1$, and
$\Omega=Y_{1,2}Y_{4,5}$, we have
\eqn\canon{\eqalign{
\ket{\pi_f}=\epsfxsize=2.cm\vcenter{\hbox{\epsfbox{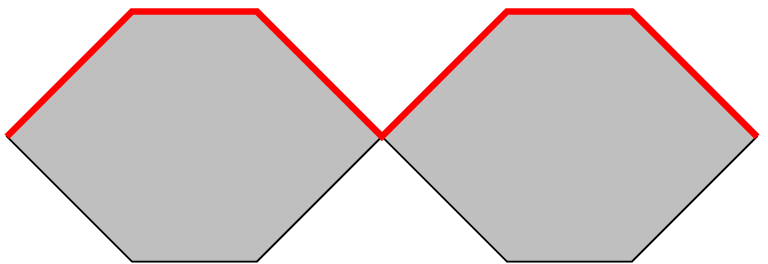}}}&= \Omega\cr
\ket{\pi_3}=\epsfxsize=2.cm\vcenter{\hbox{\epsfbox{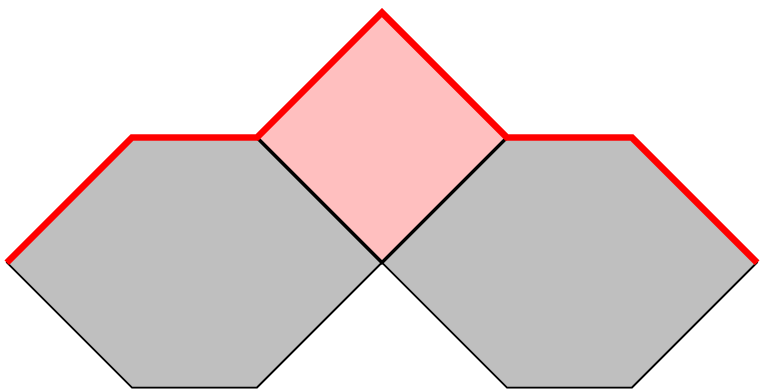}}}&= e_3 \Omega\cr
\ket{\pi_2}=\epsfxsize=2.cm\vcenter{\hbox{\epsfbox{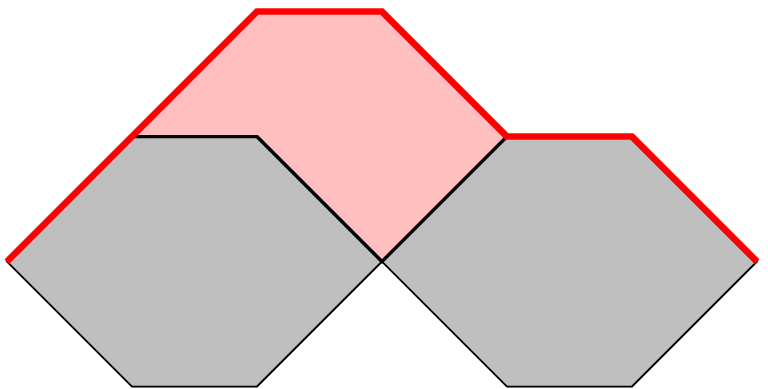}}}&= Z_{2,3} \Omega\cr
\ket{\pi_1}=\epsfxsize=2.cm\vcenter{\hbox{\epsfbox{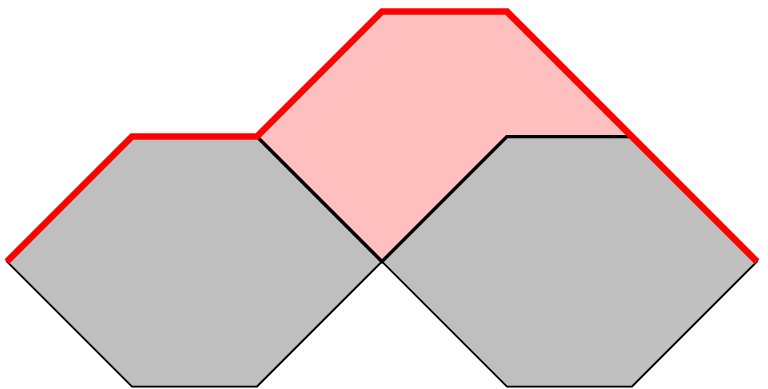}}}&= Z_{4,3} \Omega\cr
\ket{\pi_0}=\epsfxsize=2.cm\vcenter{\hbox{\epsfbox{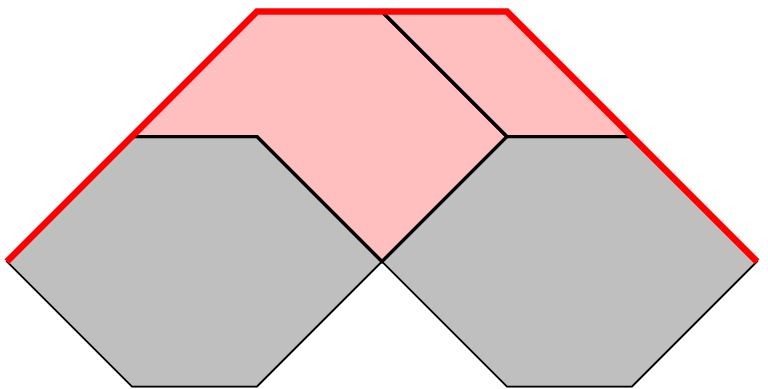}}}&= e_4 Z_{2,3} \Omega\cr
}}
and the $e_i$ read in this basis:
\eqn\actei{ 
\matrix{ 
e_1=\pmatrix{
\tau & 0  & 0 & 0 & 1\cr
0 & \tau & 1 & 0 & 0\cr
0 & 0 & 0 & 0 & 0\cr
0 & 0 & 0 & \tau & 1\cr
0 & 0 & 0 & 0 & 0} & 
e_2=\pmatrix{
\tau & 1 & 0 & 0 & 0\cr
0 & 0 & 0 & 0 & 0\cr
0 & 1 & \tau & 0 & 0\cr
0 & 0 & 0 & 0 & 0\cr
0 & 0 & 0 & 1 & \tau} & 
e_3=\pmatrix{
0 & 0 & 0 & 0 & 0\cr
1 & \tau & 0 & 0 & 0\cr
0 & 0 & \tau & 0 & 1\cr
0 & 0 & 0 & \tau & 1\cr
0 & 0 & 0 & 0 & 0} \cr 
e_4=\pmatrix{
\tau & 1 & 0 & 0 & 0\cr
0 & 0 & 0 & 0 & 0\cr
0 & 0 & 0 & 0 & 0\cr
0 & 1 & 0 & \tau & 0\cr
0 & 0 & 1 & 0 & \tau} & 
e_5=\pmatrix{
\tau & 0 & 0 & 0 & 1\cr
0 & \tau & 0 & 1 & 0\cr
0 & 0 & \tau & 0 & 1\cr
0 & 0 & 0 & 0 & 0\cr
0 & 0 & 0 & 0 & 0} \cr} }

Let us now simply list a few basic properties of this representation, of crucial importance for the following:
\item{\bf (P1)} $e_i \ket{\pi} =\tau \ket{\pi}$ 
if $\pi_i<\pi_{i+1}$ ($\pi$ locally convex)
\item{\bf (P2)} $e_i \ket{\pi}  =\sum_{\pi'} C_{i,\pi,\pi'} \ket{\pi'}$
if $\pi_i\geq \pi_{i+1}$ ($\pi$ locally flat or concave), for some 
$C_{i,\pi,\pi'}\in \{0,1\}$
\item{\bf (P3)} If $C_{i,\pi,\pi'}=1$ then $\pi'$ 
is locally convex between steps
$i$ and $i+1$, namely $\pi'_i<\pi'_{i+1}$
\item{\bf (P4)} If $C_{i,\pi,\pi'}=1$ then either $(\pi_i,\pi_{i+1})=(\pi_{i+1}',\pi_i')$
and $\pi_m=\pi'_m$ for all $m\neq i,i+1$, i.e. $\pi'$ exceeds $\pi$ by the unit lozenge 
$L_{\pi_{i+1},\pi_i}$ in the broken line representation, or
$\pi'\subset \pi$, namely the broken line representation of 
$\pi'$ lies below that of $\pi$ (or equivalently $\pi'$ may be completed into $\pi$ 
upon adding a number of lozenges).

We conclude with another important property of this representation.
There is a duality between the $SU(k)$ and $SU(n)$ representations for $N=kn$, confirmed by the 
identity of dimensions $c_{nk}^{(k)}=c_{kn}^{(n)}$.
Indeed, there exists an involution $\varphi$ from $A_{k-1}$ to $A_{n-1}$ paths of same length $kn$, defined
as follows. First of all,
$\varphi$ takes $\pi_0$ of $A_{k-1}$ to $\pi_f$ of $A_{n-1}$. The bijection is then constructed iteratively
by simultaneously subtracting (resp. adding) lozenges at the same positions from the $A_{k-1}$ (resp. 
to the $A_{n-1}$) paths, so that $\pi$ is locally convex iff $\varphi(\pi)$ is locally concave. A simpler
way of picturing this involution uses yet another expression of $A_{k-1}$ paths, as rectangular
standard Young tableaux with $k$ row of $n$ boxes, in which successive steps along the path are recorded
by their positions in the corresponding row of the tableau, i.e. if the $m$-th step is $j$,
then we write the number $m$ in the first available box from the left in the $j$-th row. 
In this language, $\varphi$ is simply the transposition of the tableaux. Moreover, this duality extends
to the path representations, namely
\eqn\idenctr{ C^{(n)}_{i,\varphi(\pi'),\varphi(\pi)}\, =\, C^{(k)}_{i,\pi,\pi'} }
In practice, this relation is extremely useful: it allows for instance to express the representation
\canon-\actei\ of $H_6^{(3)}(\tau)$ entirely in terms of the standard Dyck path representation of
$TL_6(\tau)$.

Once we pick this path representation, the $R$-matrix \trigosol\ may be interpreted as encoding 
face weights of some IRF model based on $SU(k)$, with edge degrees of freedom in the set of steps,
and acting locally at steps $i,i+1$ of the paths. This is the natural generalization of the $O(n)$
loop model in the Dyck path representation, corresponding to $k=2$ and $n=\tau$.

\subsec{qKZ equations}

\fig{The regular- (a) and boundary- (b) qKZ equations. Each oriented line carries a spectral parameter.
Each intersection between two such lines corresponds to a $\Rc$ operator. On the second line,
we have represented the diagonal operators of respectively 
multiplication of the spectral parameter by $s$ (shift), inversion of the spectral parameter
with weight $r$ ($K^{(r)}(z)$) and weight $rs$ ($K^{(rs)}(z)$), repectively pictured
as a crossed dot, empty dot, filled dot. The former is nothing but the composition 
of the two latter.}{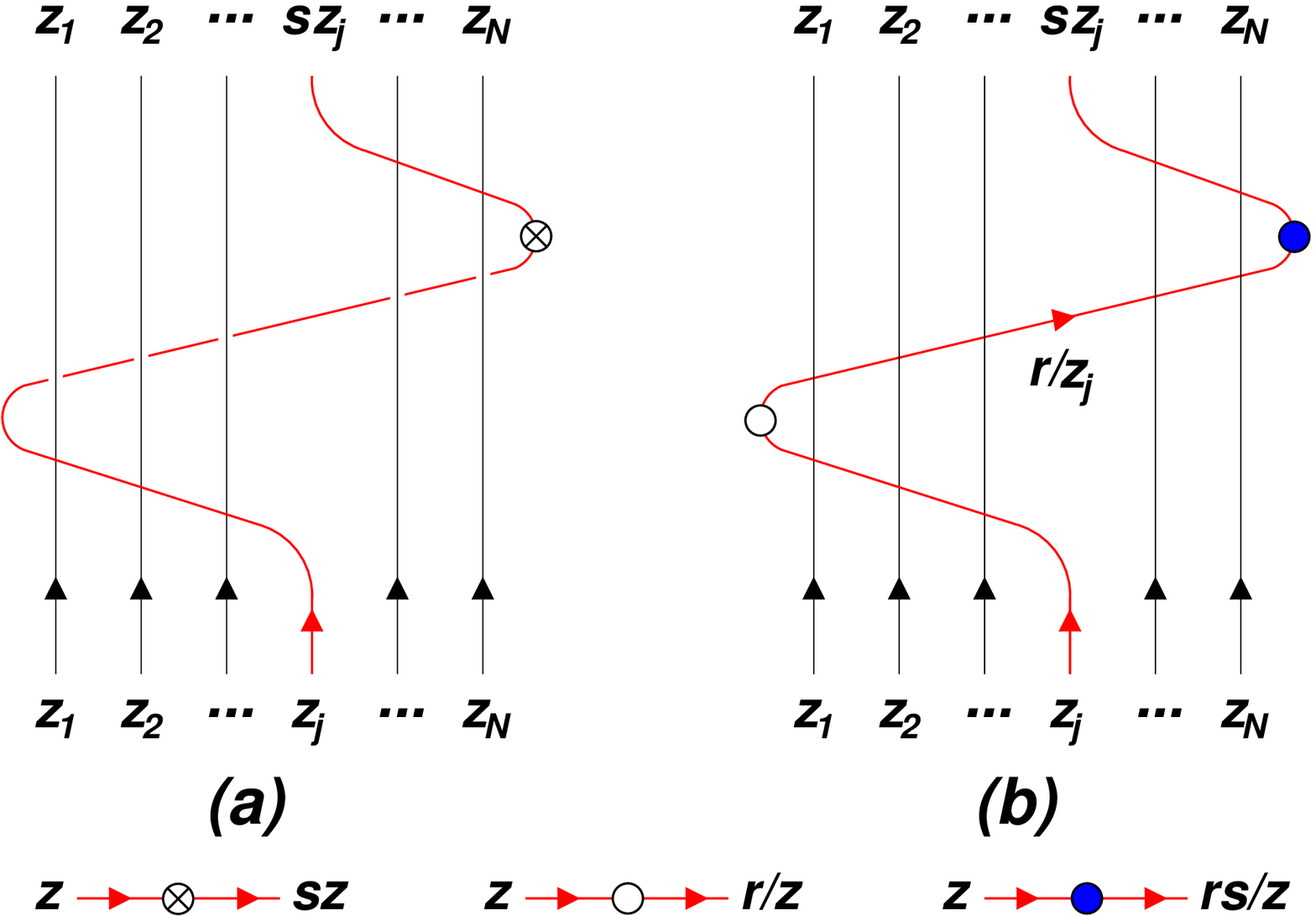}{12.cm}
\figlabel\qkzs

The qKZ equation \FR\ reads 
\eqn\qkz{\eqalign{ &\Psi(z_1,...,z_{j-1},s z_j,z_{j+1},...,z_N)=Q \Psi(z_1,...,z_{j-1},z_j,z_{j+1},...,z_N)\cr
&Q=\Rc_{j+1,j}(z_{j+1},s z_j)\ldots \Rc_{N,j}(z_N,s z_j)S_j(z)
\sigma \Rc_{1,j}(z_1,z_j)\ldots \Rc_{j-1,j}(z_{j-1},z_j)\cr}}
with the clear pictorial representation of Fig.\qkzs-(a). Here the operator $\sigma$ rotates cyclically 
the spectral parameters and the ``shift" operator $S_j(z)$ acts as the identity on the space $j$
while it multiplies the spectral parameter $z$ by some scalar $s$ (it is represented by a crossed dot
in Fig.\qkzs-(a)). 
In the pictorial representation of Fig.\qkzs-(a), each intersection between two oriented lines
stands for an $\Rc$ operator \stanpict. Note that the ``return" $z_j$ line (from left to right)
passes {\it below} the other (vertical) lines, hence there are no intersections, i.e. no $\Rc$ matrices
involved: the only effect is the operator $\sigma$, which rotates the spaces and their spectral parameters 
according to
$(z_j,z_1,...,z_{j-1},z_{j+1},...,z_N)\to (z_1,...,z_{j-1},z_{j+1},...,z_N,z_j)$.

In the presence of boundaries, the qKZ equation is modified as depicted in Fig.\qkzs-(b). The ``return line"
now crosses the vertical ones, and the shift of spectral parameter is now replaced by two boundary operators
at the far left and right of the return line. The corresponding qKZ equation now reads
\eqn\opqkz{\eqalign{  &\Psi(z_1,...,z_{j-1},s z_j,z_{j+1},...,z_N)
= {\bar Q} \Psi(z_1,...,z_{j-1},z_j,z_{j+1},...,z_N)\cr
&{\bar Q}=\Rc_{j+1,j}(z_{j+1},s z_j)\ldots \Rc_{N,j}(z_N,s z_j)K_j^{(rs)}({r\over z_j}) 
\Rc_{j,N}({r\over z_j},z_N)\ldots \Rc_{j,j+1}({r\over z_j},z_{j+1})\cr
&\ \ \times\Rc_{j,j-1}({r\over z_j},z_{j-1})\ldots  \Rc_{j,1}({r\over z_j},z_1)
K_j^{(r)}(z_j)\Rc_{1,j}(z_1,z_j)\ldots \Rc_{j-1,j}(z_{j-1},z_j)\cr}}
The boundary operator $K_j^{(r)}(z)$ acts as the identity on the space $j$, and changes the spectral parameter
$z$ into $r/z$. Together with $\Rc$, it satisfies the boundary Yang-Baxter equation of Ref.\SKLY.
In Fig.\qkzs{b}, we have represented respectively $K^{(r)}$ and $K^{(rs)}$ by an empty and a filled dot.
Here $r$ is an additional free parameter, which allows to recover in the limit $r\to 0$ the
qKZ equation without boundary \qkz\ from that with boundaries \opqkz.  Indeed, in this limit, the spectral
parameter along the return line tends to $0$: consequently, all $\Rc$
matrices along the return line degenerate into pure undercrossings, and we recover the picture of 
Fig.\qkzs-(a), upon noting that $S(z)=K^{(rs)}(r/z)K^{(r)}(z)$ is nothing but the shift operator that 
multiplies the spectral parameter $z$ by $s$.

The equation \qkz\ is equivalently standardly reduced to a system of equations for $\Psi$, namely
\eqn\sysqkz{\eqalign{
\tau_i\Psi(z_1,\ldots,z_N)&=\Rc_{i,i+1}(z_{i+1},z_i)\Psi(z_1,\ldots,z_N), \ \ i=1,2,\ldots,N-1\cr
\sigma \Psi(z_2,...,z_N,s z_1)&= c \Psi(z_1,...,z_N)\cr}}
where $\tau_i$ acts on functions of the $z$'s by interchanging $z_i$ and $z_{i+1}$, $c$ is an
irrelevant constant, and $\sigma$ now acts as a rotation of states in the representation picked for 
the $\Rc$'s.
Similarly, we may reduce \opqkz\ to the following system:
\eqna\sysopqkz{
$$\eqalignno{
\tau_i\Psi(z_1,\ldots,z_N)&=\Rc_{i,i+1}(z_{i+1},z_i)\Psi(z_1,\ldots,z_N), \ i=1,2,\ldots,N-1&\sysopqkz{a}\cr
\Psi({r\over z_1},z_2,...,z_{N-1},z_N)&= c_1(z_1) \Psi(z_1,...,z_N)&\sysopqkz{b}\cr
\Psi(z_1,z_2,...,z_{N-1},{rs\over z_N})&= c_N(z_N) \Psi(z_1,...,z_N)&\sysopqkz{c}\cr
}$$
for some functions of a single spectral parameter $c_1$ and $c_N$. Note that Eqs.\sysopqkz{b,c} 
express the covariance of $\Psi$ under the action of $K_1^{(r)}(z_1)$ and $K_N^{(rs)}(z_N)$.
The latter are obtained by showing that the operator $\bar Q$ in \opqkz\ is intertwined by $K_1^{(r)}$
and $K_N^{(rs)}$, as a direct consequence of the boundary Yang-Baxter equation.
The boundary qKZ equation \sysopqkz{} has the advantage of being ridden of the cyclicity condition 
(second line of \sysqkz),
replaced by reflective boundary conditions, somewhat easier to deal with. 
We now restrict ourselves to vectors $\Psi$ in the path representation above,
namely look for $\Psi=\sum_\pi \Psi_\pi \ket{\pi}$, $\pi$ running over the set 
of $A_{k-1}$ paths of length $N=kn$. When written in components, Eqs.\sysopqkz{} become
\eqna\compopqkz{
$$\eqalignno{
(q^{-1}z_{i+1}-qz_i)\der_i\Psi_\pi(z_1,\ldots,z_N)&=
\sum_{\pi'} C_{i,\pi',\pi} \Psi_{\pi'}(z_1,\ldots,z_N),
\ \ 1\leq i\leq N-1&\compopqkz{a}\cr
\Psi_\pi({r\over z_1},z_2,...,z_{N-1},z_N)&= c_1(z_1) \Psi_\pi(z_1,...,z_N)&\compopqkz{b}\cr
\Psi_\pi(z_1,z_2,...,z_{N-1},{rs\over z_N})&= c_N(z_N) \Psi_\pi(z_1,...,z_N)&\compopqkz{c}\cr
}$$
where the divided difference operator $\der_i$ acts on functions of the $z$'s as
\eqn\dividiff{ \der_i f= {\tau_i f -f\over z_i -z_{i+1}} }
We also denote by $t_i=(q^{-1}z_{i+1}-qz_i)\der_i$. The equations \compopqkz{a}
express nothing but $t_i\Psi=(e_i-\tau)\Psi$.

By analogy with the ordinary qKZ case, when dealing with the $A_{k-1}$ path representation
above, we write 
\eqn\levels{s=q^{2(k+l)}}
and call $l$ the level of the (boundary) qKZ equation \sysopqkz{}. 
The parameter
$r$ is free, and has the dimension of the square of a spectral parameter.
In the next section, we construct polynomial solutions of the system \sysopqkz{} for level $l=1$.

\newsec{Polynomial solutions for the level $1$ boundary qKZ equation}

We now look for polynomial solutions of level $1$ boundary qKZ
equations \sysopqkz{} with minimal degree. 

\subsec{Solutions}

We have the remarkable property for all $k$, that the equations
\compopqkz{a} may all be solved for $\Psi_\pi$ in terms solely of $\Psi_{\pi_0}$. This
relies on the property $\bf (P4)$ above of the path representation. Indeed, we may order
the paths by inclusion. Then each path $\pi$ locally convex at $i,i+1$ may be viewed
as the addition of a lozenge $L_{\pi_i,\pi_{i+1}}$ to a smaller path $\pi^-$, such that
$\pi_m^-=\pi_m$ for $m\neq i,i+1$ and $(\pi_i^-,\pi_{i+1}^-)=(\pi_{i+1},\pi_i)$. Using property $\bf (P4)$,
we see that $e_i\pi^-=\pi+\sum \pi''$, where the sum runs over paths $\pi''\subset \pi^-$.
Turning now to Eq.\compopqkz{a}, we see that, as all the $\pi'$ involved in the sum on the r.h.s.
are such that $C_{i,\pi',\pi}=1$, only one of them is strictly included in $\pi$, namely $\pi^-$, while
all others contain $\pi$. Stripping iteratively lozenges from paths, we may therefore express
each $\Psi_{\pi^-}$ as $t_i\Psi_\pi-\sum_{\pi''} \Psi_{\pi''}$, with already known $\pi''$ containing $\pi$.
This allows for expressing iteratively every $\Psi_\pi$ in terms of $\Psi_{\pi_0}$, in a triangular manner w.r.t.
inclusion of paths.

The ``fundamental" component $\Psi_{\pi_0}$ is fixed by (i) some highest weight conditions 
$(t_i+\tau)\Psi_{\pi_0}=0$
for all $i$ not multiple of $n$ (expressing that $\pi_0$ is locally flat except at the convex points
multiple of $n$, see below),  (ii) the reflective boundary conditions \compopqkz{b,c}, and finally the
requirement that it be of smallest possible degree.

As a direct consequence of Eq.\compopqkz{a}, $\Psi_\pi$ vanishes whenever $z_{i+1}=q^2 z_i$
and $\pi$ is concave or flat at $i,i+1$, namely $\pi_i\geq \pi_{i+1}$. Indeed, from property $\bf (P3)$,
the vector $\ket{\pi}$ cannot be the image of anything under $e_i$; but precisely when $z_{i+1}=q^2 z_i$,
the matrix $\Rc$ becomes proportional to $e_i$ (c.f. the definition \trigosol), hence $\Psi_\pi$ must vanish.
This is straightforwardly extended to any portion of $\pi$ flat or concave say between positions $i$ and $j$,
upon considering the appropriate product or $\Rc$ matrices: if $z_l=q^2 z_k$ for some $i\leq k<l\leq j$,
then $\Psi_\pi$ vanishes. Applying this to $\pi_0$ which is flat everywhere except
at points multiple of $n$ where it is convex, we find that $\Psi_{\pi_0}$ must
factor out the terms $\prod_{(m-1)n+1\leq i<j\leq mn} (q^2 z_i-z_j)$ for all $m=1,2,...,k$.
Alternatively, wherever $\pi_0$ is locally flat, i.e. for all $i$ not multiple of $n$,
we have $(t_i+\tau)\Psi_{\pi_0}=0$, as $\ket{\pi_0}$ has no antecedent under $e_i$: this ``highest weight
condition" is analogous to that of \LETDZJ. It implies that once the above factors are taken out of
$\Psi$, the remaining is a polynomial symmetric in the variables $\{z_{(m-1)n+1},z_{(m-1)n+2},...,z_{mn}\}$
separately for all $m=1,2,...,k$.
Finally, by an inductive argument similar to that used in Ref.\DFOP, we can show that $\Psi_{\pi_0}$ 
factors out terms $(q^{2m} r-z_kz_l)$ for all pairs $k,l$ such that
$(m-1)n+1\leq k\leq l\leq mn$, $m=1,2,...,k$. Indeed, the left boundary condition \compopqkz{b} allows
to show that $\prod_{2\leq j\leq n}(q^2 r -z_1 z_j)$ divides $\Psi_{\pi_0}$. Next, we let $z_1\to \infty$,
namely pick in $\Psi$ the coefficient of top degree in $z_1$:
in this case the space $1$ decouples from the picture of Fig.\qkzs(a), as its crossings with the 
$j$ line are undone and replaced by undercrossings, and we are left with the same boundary problem, 
but with line $1$ erased, and now a reflective
boundary condition with parameter $r$ in terms of $z_2$. Repeating iteratively this process,
namely taking successively $z_2,z_3,...,z_{n-1}\to\infty$ exhausts all the announced extra factors
for $m=1$. Assuming each variable occurs in $\Psi$ with maximal degree $2(n-1)$, we find that the 
leftover dependence on $z_n$ must be proportional to $z_n^{n-1}$, and we may write
$\Psi_{\pi_0}(\infty,...,\infty,z_n,z_{n+1},...,z_N)=C z_n^{n-1} F(z_{n+1},...,z_N)$.
The full vector $\Psi(\infty,...,\infty,z_n,z_{n+1},...,z_N)$ now obeys a left reflective boundary
condition of the form 
\eqn\newcond{\Psi(\infty,...,\infty,{r\over z_n},z_{n+1},...,z_N)
=c_1(z_n)\Psi(\infty,...,\infty,z_n,z_{n+1},...,z_N)}
We now note that $\pi_0=e_n\pi^-$, for the unique $\pi^-$ such that $\pi_i^-=\pi_i$ for $i\neq n,n+1$, and 
$(\pi_n^-,\pi_{n+1}^-)=(\pi_{n+1},\pi_n)$. According to \compopqkz{a}, 
we have $t_n\Psi_{\pi_0}=\Psi_{\pi^-}$. Expressing that the component $\Psi_{\pi^-}$ obeys
the new left reflection condition \newcond\ solely in terms of $\Psi_{\pi_0}$ yields
the following relation for $F$:
\eqn\relF{\eqalign{ c_1(z_n)\,{q^{-1}z_{n+1}-q z_n\over z_n-z_{n+1}}&
\big((z_{n+1})^{n-1}F(z_n,...)-(z_n)^{n-1}F(z_{n+1},...)\big)\cr
&={q^{-1}z_nz_{n+1}-qr\over qr-z_nz_{n+1}}\big((z_{n+1})^{n-1}
F({r \over z_n},...)-\Big({r\over z_n}\Big)^{n-1}F(z_{n+1},...)\big)\cr}}
Picking $z_n=q^{-2}z_{n+1}$ finally yields $F({q^2r \over z_{n+1}},...)=(q^2r/z_{n+1}^2)^{n-1}F(z_{n+1},...)$
and therefore $\Psi_{\pi_0}$ now has all the announced extra factors for $m=2$ (everything works now as if the
left boundary parameter had been rescaled as $r\to rq^2$). Proceeding iteratively,
we finally get all the announced factors for $m=1,2,...,k$. We claim that the solution of minimal degree
has no extra factors in its $\Psi_{\pi_0}$, namely that
\eqn\psizer{\Psi_{\pi_0}=\prod_{m=1}^k \prod_{(m-1)n+1\leq i<j\leq mn} (q^2 z_i-z_j)(rq^{2m} -z_iz_j)}
and that Eqs.\compopqkz{} are satisfied for $s=q^{2(k+1)}$ and
\eqn\constants{ c_1(z)=\left({r\over z^2}\right)^{n-1},\qquad c_N(z)=\left({r q^{2(k+1)}\over z^2}\right)^{n-1}}
This fixes the solution entirely.  

As in the case without boundaries of \LETDZJ, the solution may alternatively be fixed by 
some global requirement that it vanishes for some specific choices of the spectral parameters.
It was found indeed that in the case without boundary $\Psi$ must vanish for any ordered sequence
of $k+1$ spectral parameters of the form $\{z,q^2 z,...,q^{2k}z\}$, the so-called ``quantum incompressibility"
condition, first noticed by Pasquier \Pas\ in the case $k=2$, and allowing for interpreting
$\Psi$ as $q$-deformed wave function of fractional quantum Hall effect.
In the present case with boundaries, we find that
$\Psi$ must vanish for any ordered $k+1$-tuple of spectral parameters of the form
\eqn\ktuples{\{z,q^2 z,...,q^{2(m-1)}z,{r q^2\over z},{r q^4\over z},...,{rq^{2(k-m)}\over z}\}}
for $m=2,3,...,k+1$. This generalizes the quantum incompressibility condition, in that
the wave function must vanish whenever $k+1$ electrons come into contact with one 
another or now with their {\it reflections}.

As in the case of \LETDZJ, we may also derive recursion relations for the components, relating size
$N=kn$ to size $N'=k(n-1)$. This is done by simply taking $z_1=z,z_2=q^2 z,...,z_k=q^{2(k-1)}z$,
in which case only paths with $\pi_i=i$ for $i=1,2,...,k$ have non-vanishing components in $\Psi$.
The rest of the path, $\pi'=(\pi_{k+1},...,\pi_N)$ is an $A_{k-1}$ path of length $N'=N-k$,
and we have
\eqn\recuop{ \Psi_{\pi}(z_1,...,z_N)\bigg\vert_{z_1=q^{-2}z_{2}=...=q^{-2(k-1)}z_k=z}=
\left(\prod_{j=k+1}^N  (q^{2k}z-z_j)(rq^2-z z_j)\right) \Psi_{\pi'}(z_{k+1},...,z_N) }
where the prefactor is fixed by the degree and vanishing requirements.

\subsec{Other sizes}

As already touched-upon in the previous section, there is an easy way of extending our polynomial
solutions to all sizes $N$, not necessarily a multiple of $k$. Starting from a solution at $N=kn$,
we may indeed send successively the first $j$ spectral parameters to $0$
(or equivalently to $\infty$), leaving us with only non-vanishing
components of $\Psi$ for paths $\pi$ starting with the (unique) convex sequence $1,2,3,...,j$.
This provides us with a solution to the boundary qKZ equation of $z_{j+1},...,z_N$, hence in
size $N-j$. Note that now in the corresponding representation
the paths start at the point $u_1+u_2+...+u_j$ in the Weyl chamber of $SU(k)$ and end at its origin.
This construction is valid for all $j=1,2,...,k-1$ hence allows to exhaust all possible sizes.

\newsec{Sum rules}

\subsec{Generalizations of the Razumov--Stroganov sum rule at $q=-e^{i\pi\over k+1}$}

We proceed as in \LETDZJ. We introduce a covector $v$ such that
\eqn\covec{ e_iv=\tau v, \qquad i=1,2,...,N-1}
This covector is unique under the condition that all its entries be strictly positive
and that $v_{\pi_f}=1$, and exists only if $q=-e^{i\pi\over k+1}$ or its conjugate,
namely $\tau=2\cos{\pi\over k+1}$. The entries of $v$ are directly constructed from the path
representation and may be expressed as products of ratios of the Chebyshev polynomials $U_m(\tau)$ with $m<k$,
associated to the tiles in the decomposition of the paths, and
all of which are positive for the above choice of $\tau$. For illustration, for $k=3$, $n=2$, we
take $\tau=\sqrt{2}$, and 
associate the factor $U_1(\tau)=\sqrt{2}$ to every lozenge and 
the factor $U_2(\tau)=1$ to every tile made of two lozenges
in the path decomposition \canon, corresponding
respectively to the left eigenvalues of $v$ on $e_i$ and on $Z_{i,j}$, resulting in
$v=\{1,\sqrt{2},1,1,\sqrt{2}\}$ in the path basis.

The condition \covec\ immediately implies that $v\Rc_{i,i+1}(z,w)=v$, readily checked from the definition
\trigosol. Henceforth, the combination $v\cdot \Psi$ is a symmetric polynomial of the $z$'s,
as a direct consequence of Eq.\sysopqkz{a}.
Introducing 
\eqn\ikopdef{I_{kn}^{(k)}(z_1,z_2,...,z_{kn}\vert r)\equiv (-i/q)^{kn(n-1)/2}v\cdot \Psi(z_1,...,z_{kn})}
with $i=\sqrt{-1}$,
we find that $I_{kn}^{(k)}$ is a polynomial of degree $kn(n-1)/2$ of $r$, whose coefficients
are all symmetric polynomials of the $z$'s, and the top degree and constant coefficients
are respectively 
\eqn\topbo{\eqalign{I_{kn}^{(k)}\big\vert_{r^{kn(n-1)/2}}&= s_Y(z_1,...,z_{kn})\cr
I_{kn}^{(k)}\big\vert_{r^0}&= (z_1z_2...z_{kn})^{n-1}\, s_Y(z_1,...,z_{kn})\cr}}
where $s_Y$ is the Schur function associated to the Young diagram $Y$ with $k$ rows of $n-1$ boxes,
$k$ rows of $n-2$ boxes, ..., $k$ rows of $1$ box. The second line of \topbo\ is nothing but
the generalized Razumov-Stroganov sum rule of \LETDZJ, recovered in the limit $r\to 0$ as explained above.
The polynomial $I_{kn}^{(k)}(z_1,...,z_{kn}\vert r)$ has only non-negative integer coefficients 
in $r$ and the $z$'s. 
Note also the covariance under $z_i\to r/z_i$ for each $i$, inherited from the boundary reflections
\sysopqkz{b,c}\ and the overall symmetry, namely
\eqn\covaI{ I_{kn}^{(k)}(z_1,...,z_{j-1},{r\over z_j},z_{j+1},...,z_{kn}\vert r)=\left({r\over z_j^2}\right)^{n-1}
I_{kn}^{(k)}(z_1,...,z_{j-1},z_j,z_{j+1},...,z_{kn}\vert r) }
which allows to relate the coefficients of $I_{kn}^{(k)}$ for different powers of $r$. In particular,
taking simultaneously all $z_i\to r/z_i$ relates the top and constant coefficients, and the
first line of \topbo\ follows from the second.
The polynomial $I_{kn}^{(k)}$ \ikopdef\ is entirely determined up to global normalization
by the following properties:
\item{(i)} it is a symmetric polynomial of the $z$'s quasi-homogeneous under
$z_i\to \lambda z_i$ for all $i$ and $r\to \lambda^2 r$, with total degree $3kn(n-1)/2$ and partial
degree $2(n-1)$ in each of the $z$'s and $kn(n-1)/2$ in $r$
\item{(ii)} it vanishes for all the choices of ordered $k$-tuples of spectral parameters \ktuples, 
with $q=-e^{i\pi\over k+1}$
\par
\noindent and the normalization is fixed by \topbo.

Let us first discuss the Temperley-Lieb case $k=2$ and $n$ arbitrary, where drastic simplifications occur. 
At the Razumov--Stroganov point
$q=-e^{i\pi\over 3}$, $\tau=1$, we have $v_\pi=1$ for all $\pi$, hence $I_{2n}^{(2)}$ is simply the sum
of components of $\Psi$. The quantity $I_{2n}^{(2)}$ reduces to the result
of \DFOP\ at $r=1$. For generic $r$, the determinantal formulas of \DFOP\
are straightforwardly generalized to
\eqn\ikoptwo{\eqalign{ I_{2n}^{(2)}(z_1,...,z_{2n}\vert r)&=
{\prod_{i,j=1}^n(z_i^2+z_iz_{j+n}+z_{j+n}^2)(r^2+rz_iz_{j+n}+z_i^2z_{j+n}^2)\over
\prod_{1\leq i<j\leq n} (z_i-z_j)(r-z_iz_j)(z_{i+n}-z_{j+n})(r-z_{i+n}z_{j+n})}\cr
&\ \ \ \times \det_{1\leq i,j\leq n}\left( {1\over z_i^2+z_iz_{j+n}+z_{j+n}^2}
{1\over r^2+rz_iz_{j+n}+z_i^2z_{j+n}^2}\right)\cr}}
which is an open version of the Izergin--Korepin determinant \IK,
and the Pfaffian formula becomes
\eqn\ikopfatwo{\eqalign{ I_{2n}^{(2)}(z_1,...,z_{2n}\vert r)^2&=
\prod_{1\leq i<j\leq 2n}^n{(z_i^2+z_iz_{j}+z_{j}^2)(r^2+rz_iz_{j}+z_i^2z_{j}^2)\over
(z_i-z_j)(r-z_iz_j)}\cr
&\ \ \ \times {\rm Pf}_{1\leq i<j\leq 2n}\left( 
{(z_i-z_j)(r-z_iz_j)\over (z_i^2+z_iz_{j}+z_{j}^2)(r^2+rz_iz_{j}+z_i^2z_{j}^2)}\right)\cr}}
These are proved for instance by using the symmetry and the recursion relations
obeyed by $v\cdot \Psi$ as a consequence of Eq.\recuop, and showing that the
r.h.s. is the unique symmetric polynomial obeying the same constraints.
In the homogeneous case where all $z_i=1$, $I_{2n}^{(2)}$ reduces to a reciprocal polynomial of
$r$ with non-negative integer coefficients, equal respectively to $3^{n(n-1)/2}$ times
the number $A_n$ of alternating sign matrices (ASM) of
size $n\times n$ at $r=0$ (a result proved in \DFZJ),
and to $3^{n(n-1)}$ times the number $A_V(2n+1)$
of vertically symmetric alternating sign matrices (VSASM) of size $(2n+1)\times (2n+1)$
at $r=1$ (a result proved in \DFOP), with respective values
\eqn\vsasm{ A_n=\prod_{j=0}^{n-1} {(3j+1)!j!\over (2j+1)!(2j)!},\qquad {\rm and}\qquad
A_V(2n+1)=\prod_{j=0}^{n-1} {(6j+4)!(2j+2)!\over (4j+4)!(4j+2)! } }

For general $k$, we have the following expression for the sum rule $I_{kn}^{(k)}$:
\eqn\genefor{ I_{nk}^{(k)}(z_1,...,z_{nk}\vert r)={\tilde s}_Y(z_1,...,z_{nk}\vert r) }
where ${\tilde s}_Y$ are ``symplectic" Schur functions depending on the extra parameter $r$,
and defined as
\eqn\defsimplecshur{ {\tilde s}_Y (z_1,...,z_{nk}\vert r)
=z_1z_2...z_{kn} \ {\det_{1\leq i,j\leq kn} \left( z_i^{\ell_{kn+1-j}+j}-(r/z_i)^{\ell_{kn+1-j}+j}\right) \over
\det_{1\leq i,j\leq kn} \left(z_i^{j}-(r/z_i)^j\right) }}
with $Y$ as in \topbo, and $\ell_i=n-1-[(i-1)/k]$, is the number of boxes in the $i$-th row,
$i=1,2,...,kn$ (we have added $k$ bottom rows of $0$ box for convenience).
The sum rule \genefor\ is proved as usual by checking that the r.h.s. fulfills all requirements
of degree and vanishings above, and is therefore fixed by uniqueness. This formula is
a nice multi-parameter generalization of that written by Okada \OKA\ for VSASMs, 
in terms of specialized characters of the symplectic group.

In the particular case $n=2$ and $k$ arbitrary, It is easy to show that \genefor\ reduces to
\eqn\exatwo{ I_{2k}^{(k)}(z_1,...,z_{2k}\vert r)= \sum_{m=0}^k r^m s_{Y_m}(z_1,...,z_{2k})}
expressed in terms of ordinary Schur functions for the young diagrams $Y_m$
with $k-m$ rows of $2$ boxes and $k$ rows of $1$ box.
Consequently, in the homogeneous limit where all $z_i\to 1$, we have
\eqn\valtwoi{ I_{2k}^{(k)}(1,1,...,1\vert r)=\sum_{m=0}^k r^m {k+1\over 2k+1}{2k+1\choose m}{2k+1\choose k-m}}
and we get
\eqn\recov{A_2^{(k)}={1\over k+1}I_{2k}^{(k)}(1,1,...,1\vert 0)=c_{k} }
the $k$-th Catalan number, while at $r=1$ we obtain 
\eqn\aval{A_V^{(k)}(2)={1\over (k+1)^2}I_{2k}^{(k)}(1,1,...,1\vert 1)=2 {(4k+1)!\over (3k+2)! (k+1)!}}

$$\vbox{\font\bidon=cmr8 \def\bidondon{\bidon} \bidondon \offinterlineskip
\halign{\tv \quad # \tv & 
\hfill \ # \hfill & \hfill # \hfill & \hfill # \hfill &  \hfill # \hfill 
& \hfill # \hfill \tv \cr
\noalign{\hrule}
\tvi  $\scriptstyle n\backslash k$&\hfill  1 \hfill &\hfill  2 \hfill 
&\hfill  3 \hfill &\hfill  4 \hfill &\hfill  5 \hfill \cr
\noalign{\hrule}
\tvi 1 & 1 & 1 & 1 & 1 & 1 \cr
\tvi 2 &  1 & 3  & 13  & 68  & 399  \cr
\tvi 3 &  1 &  26 & 1938  &  246675 &  43475640 \cr
\tvi 4 &  1 &  646 &  3251625 & 46278146640  &  1238084585216726 \cr
\tvi 5 &  1 &  45885 & 61003011480  &  444316978958627636 &  9080679253196247653297250 \cr
\noalign{\hrule} 
}}$$
\noindent{\bf Table I:} The numbers $A_V^{(k)}(n)$ for $1\leq n,k\leq 5$.
\vskip .5cm

For arbitrary $k$, the homogeneous case of all $z_i=1$ yields again for
$I_{kn}^{(k)}$ a reciprocal polynomial of $r$, with non-negative integer coefficients. 
The constant coefficient of this polynomial
is the result of the homogeneous limit of the generalized
Razumov-Stroganov sum rule computed in \LETDZJ, and reads $(k+1)^{n(n-1)/2}$ times
an integer $A_n^{(k)}$ generalizing the number of ASMs $A_n=A_n^{(2)}$, 
and which may be rewritten as
\eqn\genasm{ A_n^{(k)}=\prod_{j=0}^{n-1}{j!\prod_{i=1}^{k-1}\big((k+1)j+i\big)!\over 
\prod_{i=0}^{k-1}(kj+i)! }}
The numbers \genasm\ are nothing but the properly normalized
dimensions of the $GL_{kn}$ representations $Y$ \FULHA.
At $r=1$ the homogeneous value of $I_{kn}^{(k)}$ reads
$(k+1)^{n(n-1)}$ times a number $A_V^{(k)}(n)$
generalizing the number of $(2n+1)\times (2n+1)$ VSASMs 
$A_V(2n+1)$. We find that
\eqn\genav{A_V^{(k)}(n)=\prod_{j=1}^{n-1} {(2j)!(2nk+2j-1)!\over ((k+1)j)!((k+1)(j+n)-1)!}}
These numbers are the properly normalized
dimensions of the $SP_{2kn}$ representations $Y$ \FULHA.
The numbers $A_V^{(k)}(n)$ are easily checked to reduce to $A_V(2n+1)$ for $k=2$ and all $n$,
and to \aval\ for $n=2$, and all $k$. The numbers $A_V^{(k)}(n)$
are displayed in Table I for $n,k=1,2,...,5$. For illustration, for $k=3$, $n=2$ we have
$I_6^{(3)}(1,1,...,1\vert r)=20+84r+84r^2+20r^3$, from which we read the values of
$A_2^{(3)}=20/4=5$ and $A_V^{(3)}(2)=(20+84+84+20)/4^2=13$.

As mentioned in Sect.3.2, we have access to analogous sum rules for arbitrary sizes $N$ not necessarily
a multiple of $k$ by sending spectral parameters to $0$ and reducing the space of path states 
accordingly. In size $nk-j$, the sum rule simply reads \genefor, with say 
$z_{1}=z_{2}=...=z_{j}=0$ and where the first $j$ rows have been deleted from $Y$. 
This leads to more integer numbers in the homogeneous limit.

\subsec{More combinatorics at $q=-1$}

In \LETDZJ, a connection was established between the ``rational limit" $q\to -1, z_i\to 1$ of the 
$U_q(\frak{sl}(k))$ level 1 minimal polynomial solution of the qKZ equation without boundary
and so-called extended Joseph polynomials, related to the geometry of the
variety of upper triangular matrices with vanishing $k$-th power. 
As a particular
corollary of this relation, the entries of $\Psi$ were found to tend to positive integers
in the rational homogeneous limit, in which one lets first all $z_i\to 1$, and then $q\to -1$
after dividing out by the appropriate power of $q+1$. These integers were interpreted as
the degrees of the components of the abovementioned variety.

In the present case, we may also consider an analogous rational limit. 
Of course, at $r=0$, we may reproduce the rational limit of \LETDZJ.
If we want a different 
result from that of \LETDZJ, we need to let $r\to 1$ as well.
Substituting
\eqn\scalim{ q=-e^{-\epsilon a/2}, \qquad z_i=e^{-\epsilon w_i}, \qquad r=e^{-\epsilon \rho} }
into our minimal polynomial solution $\Psi$ of the $U_q(\frak{sl}(k))$ level 1 boundary
qKZ equation, and taking $\epsilon\to 0$, we obtain a two-parameter ($a,\rho$) family of polynomials
of the $w$'s. At $a=\rho=0$, these are some sorts of reflected versions
of the Joseph polynomials \Jo, which await some algebro-geometric interpretation.
A remarkable property is that for $a=1,\rho=0$, and in the homogeneous limit
where $w_i\to 0$ for all $i$, the above polynomials reduce to {\it non-negative integers}, 
yet to be interpreted as degrees of components of some matrix variety.

Remarkably, for $k=2$ and $n$ arbitrary, we have found that for a suitable normalization of $\Psi$
these would-be degrees sum to the total degree of the ``Brauer scheme" considered in \KZJ, suggesting a
possible connection. More precisely, taking $r=1$ and $z_i=1$ for all $i$ in our solution $\Psi$, we have
\eqn\morpre{ \lim_{q\to -1} {1\over (q^2-1)^{2n(n-1)}} \sum_\pi \Psi_\pi(1,1,...,1) =
\det_{0\leq i,j \leq n-1} {2i+2j+1\choose 2i} }
where the numbers on the r.h.s. are the degree of the Brauer scheme
$\{ M\in M_{2n}(\IC), M \bullet M=0 \}$  where $\bullet$ stands for a suitable deformation
of the ordinary matrix product \KZJ, also obtained as sum rules in the O(1) crossing loop model
on a semi-infinite cylinder \DFZJb.
For illustration, in the cases $n=1,2,3$ we find the following limits $\psi_{2n}^{(1)}$ 
when $q\to -1$ of $\Psi(z_i=1\vert r=1)/(q^2-1)^{2n(n-1)}$:
\eqn\illu{\psi_2^{(1)}=\{1\},\quad 
\psi_4^{(1)}=\{5,2\},\quad 
\psi_6^{(1)}=\{149, 52, 58, 40, 8\}}
expressed in the standard Dyck path basis of $TL_6(\tau)$, ordered by inclusion.
The components of \illu\ sum respectively to $1,7,307$, the degrees of the Brauer scheme for matrices of size
$2,4,6$ respectively. For completeness, the corresponding $r=0$ solutions of \LETDZJ\ yield
the following limits $\psi_{2n}^{(0)}$ when $q\to -1$ of $\Psi(z_i=1\vert r=0)/(q^2-1)^{n(n-1)}$:
\eqn\illulim{\psi_2^{(0)}=\{1\},\quad 
\psi_4^{(0)}=\{2,1\},\quad 
\psi_6^{(0)}=\{10, 4, 4, 4, 1\}}
The above observation is particularly intriguing, as we already know that another particular point
($q=-e^{i\pi\over 3}$) of the {\it same} solution $\Psi$ also displays entries
leading to non-negative integers for both $r=0$ and $r=1$, as mentioned in the previous section. 
These entries $\Psi_\pi$ were actually identified
via the full Razumov--Stroganov conjecture \RS\ to fully-packed loop configurations on a $n\times n$
square grid (at $r=0$) or to vertically symmetric fully-packed loop configurations on a 
$(2n+1)\times (2n+1)$ square grid (at $r=1$) connected in both cases
according to the same link pattern as that encoded in the Dyck path $\pi$. 

The case $k>2$ also seems to lead to non-negative integer entries in the rational homogeneous limit, 
provided we take $r=1$ and all $z_i=1$, and then consider
\eqn\limk{ \lim_{q\to -1} {1\over (q^2-1)^{kn(n-1)}}\sum_\pi \Psi_\pi(1,1,...,1) }
These are presumably
related to the degrees of higher order generalizations of the Brauer loop scheme \DFKZJ.
For illustration, we have for $k=3$ and $n=2$, in the basis \canon:
\eqn\illuthree{ \psi_{6}^{(1)}=\{ 60, 28, 13, 15, 6\} }
whereas the corresponding $r=0$ limit (with $(q^2-1)^{kn(n-1)}$ replaced by $(q^2-1)^{kn(n-1)/2}$
in the limit of \limk) yields in the basis \canon:
\eqn\illuclo{ \psi_{6}^{(0)}=\{6, 3, 2, 2, 1 \}}

As mentioned in Sect.3.2, we have access to analogous integer $\Psi$'s 
and sum rules for arbitrary sizes $N$ not necessarily
a multiple of $k$, presumably relating to generalized Brauer schemes of 
matrices of corresponding sizes.

\newsec{Conclusion}

In this note, we have constructed the minimal
polynomial solutions to the level 1 $U_q(\frak{sl}(k))$ boundary qKZ equation, 
and investigated its ``combinatorial points".
We have derived some
higher rank generalizations of the Razumov-Stroganov sum rules for the open boundary O(1)
spin chain. We have actually found a family of solutions depending on one parameter $r$
playing the role of boundary magnetic field.  The closed and open boundary cases of the
Razumov--Stroganov sum rules are recovered for $r=0$ and $1$ respectively.
Sum rules for the components of the solution \`a la Razumov--Stroganov are generically obtained at 
a unique point $\tau=2\cos{\pi\over k+1}$ for each $k$, where a covector $v$ with the properties
\covec\ and with positive entries can be constructed. 

In view of trying to cook up a full conjecture \`a la Razumov--Stroganov, we may try to
consider the vector $w$ with entries $w_\pi=(-i/q)^{kn(n-1)/2}v_\pi \Psi_\pi$: 
unfortunately, these entries
are not real for generic $r$. For illustration, for $k=3$, $n=2$, the corresponding entries read 
in the basis \canon:
\eqn\vecillu{\eqalign{w_f=4(1 + &9r + 9r^2 + r^3), w_3=4(1 + 7 r + 7r^2 + r^3),
w_2=4(1 + (2 +i) r + (2 - i) r^2 +r^3),\cr
w_1&=4(1 + (2 -i) r + (2 + i) r^2 + r^3),\qquad w_0= 4( 1 + r + r^2 + r^3)\cr} }
and sum to the abovementioned value of $I_6^{(3)}(1,1,...,1\vert r)$.
In this particular case however, the entries of $w$ reduce to {\it non-negative integers} 
in both cases $r=0,1$, with $w^{(0)}=4\{1,1,1,1,1\}$ and $w^{(1)}=8\{10,8,3,3,2\}$. 
In general, it is tempting to conjecture that $w$ has integer entries, and we have
indeed observed this fact for $r=0$ and the cases $n=2$, $k=3,4,5$, and $n=3$, $k=3$.
For $r=1$ however we have found a counterexample for $n=2$, $k=4$, in which the $14$ entries
of $w$, although positive and summing to $4^2A_V^{(4)}(2)=16\times 68$, are not all integers. 
So if a full Razumov--Stroganov conjecture exists it has to be more subtle in the $r=1$ case.

Note that, as we have $q^{2(k+1)}=1$ at the generalized
Razumov--Stroganov point, the left and right boundary conditions in \sysopqkz{} coincide.
When $r=0$, we are precisely
in the case where $\Psi$ is the groundstate vector of the inhomogeneous
$A_{k-1}$ IRF model on a semi-infinite cylinder of perimeter $N$ as pointed out in \LETDZJ,
and governed by the cyclic Hamiltonian $H=\sum_{1\leq i\leq N} (\tau-e_i)$, 
$e_N=\sigma e_{N-1}\sigma^{-1}$ an additional
generator for the cyclic Hecke algebra.
For $r=1$, $\Psi$ is the groundstate vector of the inhomogeneous
$A_{k-1}$ IRF model on a semi-infinite strip of width $N$, with reflecting boundary
conditions, and governed by the Hamiltonian $H=\sum_{1\leq i\leq N-1} (\tau-e_i)$. 
In both cases, the entries $(-i/q)^{kn(n-1)/2}\Psi_\pi$ and $v_\pi$ are manifestly positive, 
as these are respectively
the entries of the right and left Perron-Frobenius eigenvectors of the corresponding Hamiltonians.
Alternatively, we may view 
the $r=1$ model as describing a stochastically ``growing interface" like in \PRdGN\ for 
the Temperley--Lieb case $k=2$, upon interpreting the action of 
$\Rc_{i,i+1}(z_i,z_{j})$ \trigosol\
on a path as the identity with probability $1-t_{i,j}=(z_j -q^2 z_i)/(q^2 z_j-z_i)$ and as $e_i/\tau$ 
with probability $t_{i,j}$, which (i) leaves the path unchanged (if $\pi$ is convex at $i,i+1$),
(ii) adds a lozenge to it (if $\pi$ is concave at $i,i+1$) (iii) or shrinks it to a smaller path $\pi'$
such that $C_{i,\pi,\pi'}=1$ (if $\pi$ is flat or concave at $i,i+1$), the last two operations
weighted by $1/\tau$. The combination of entries 
${\cal P}(\pi)=v_\pi\Psi_\pi/(v\cdot \Psi)$ may then be interpreted 
as the {\it invariant probability} for the interface to be equal to the path $\pi$. 
These invariant probabilities read in the case $k=3$, $n=2$: 
$\{{5\over 13},{4\over 13},{3\over 26},{3\over 26},{1\over 13}\}$ for the paths of \canon.
The above counterexample to integrality simply shows that these invariant probabilities
will not be rational numbers in general. It is still possible nevertheless that generalizations
of the observables considered in \PRdGN\ could lead to simple rational expectation values. 
For instance, from our numerical data, we conjecture that the probability
of convex transitions $\pi_i<\pi_{i+1}$ in the above interfaces reads
\eqn\conjec{ C(2,k)={(k-1)(13k+4)\over 2(2k-1)(4k+1)}}
for the $A_{k-1}$ growth problem with $n=2$. This is easily checked for $k=3$, as the paths
of \canon\ have respectively $\{4,3,3,3,2\}$ convex transitions among their $5$
transitions, leading to a probability
${1\over 5}\{4,3,3,3,2\}\cdot\{{5\over 13},{4\over 13},{3\over 26},{3\over 26},{1\over 13}\}={43\over 65}$.

The sum rules obtained in this paper produce nice integer sequences, which we expect to be counting
generalizations of ASMs and of VSASMs. A first step in understanding what exactly is counted by these
numbers could be to find some higher $k>2$ analogues of the Izergin--Korepin determinant (or its open
version \ikoptwo) for our sum rules $I_{kn}^{(k)}(z_1,...,z_{kn}\vert r)$. We may also try to generalize 
the Pfaffian formula \ikopfatwo.  
The hope would be eventually
to be able to define an $A_{k-1}$ vertex model with a particular type of (domain-wall)
boundary conditions on a finite domain, whose partition function exactly matches 
$I_{kn}^{(k)}(z_1,...,z_{kn}\vert r)$. Then the configurations of such a model would form the desired
generalizations of ASMs or VSASMs. This might then open the route to generalizations of other
symmetry classes of such objects, according to the symmetries of the above domain, in the spirit of
\KUP.

As a side remark, the number of different objects enumerated by the ASM number $A_n$ 
is quite considerable: totally symmetric self-complementary plane partitions (TSSCPP),
descending plane partitions (DPP), rhombus tilings of particular domains of the triangular 
lattice, osculating walks on a square grid, ice configurations
on a square with domain wall boundaries etc, and these give rise to many different
formulas for $A_n$, as determinants or Pfaffians.  We may wonder which ones of these objects and/or
formulas will be amenable to higher $k$ generalizations, maybe all of them?

The other combinatorial point $q=-1$ ($\tau=2$) seems to be of a different nature. It corresponds to 
the rational limit of the above $A_{k-1}$ IRF models, namely giving rise to a rational 
solution of the Yang-Baxter equation. Another rational solution of a similar nature was considered in
\DFZJb\ and led to sum rules for the O(1) model of crossing loops. All these models seem to be related 
to the geometry of some specific matrix varieties, allowing in 
particular for interpreting in the homogeneous limit
the various integer entries of $\Psi$ at hand as degrees of the components
of these varieties. The identification of these varieties for the open boundary 
cases is under way \DFKZJ.

\medskip

\noindent{\bf Acknowledgments:}  We thank M. Bauer, P. Zinn-Justin and J.-B. Zuber for useful discussions.
This work is partly supported by the Geocomp project (ACI {\it Masse de Donn\'ees}) and
the European network ``ENIGMA", grant MRT-CT-2004-5652. 

\listrefs
\end